# Electron multipacting in long-bunch beam

Li Kai-Wei(李开玮)[1]

Institute of high energy physics, Chinese Academy of Sciences, Beijing 100049, China

Guang Dong Polytechnic College, Zhaoqing 526100, China

**Abstract**

The electron multipacting is an important factor for the development of the electron cloud. There is a trailing-edge multipacting in the tail of the long-bunch beam. It can be described by the energy gain and motion of electrons. The analyses are in agreement with the simulation.

**Key words:** trailing-edge multipacting, electron cloud, long-bunch beam.

## I. Introduction

The electron cloud instability has long persisted. In 1971, an electron cloud caused beam instability was observed at CERN Intersecting Storage Rings during a coasting-beam operation and was cured with clearing electrodes [1, 2]. After that it was found in many proton circular accelerators, such as LANL PSR [3], CERN SPS [4] and SNS in ORNL [5]. The beam induced electron multipacting is responsible for the development of the electron cloud. In long-bunch beam, the single-bunch, trailing-edge multipacting is probably dominant. The mechanism of multipacting can be explained by the electron motion and energy gain. Generally, a proton bunch has a mountain-like shape in the longitudinal coordinate. When the leading edge of a bunch passes through an electron cloud region, the electrons in the vacuum chamber are trapped and made to oscillate inside the proton bunch. These electrons consist of carryover electrons that survived the beam gap and primary electrons produced by lost protons hitting the chamber surface. After the peak of bunch passes, these electrons are released from the proton bunch, hit the surface of the vacuum chamber, and cause secondary emission. Furthermore, the primary electrons produced in the trailing edge can typically gain enough energy from the beam to cause secondary emission. The paper is organized as follows. First, we introduce the simulation model. Second, we explore the motion of electron in the long-bunch beam. Third, the simulation result is presented. At last, the conclusion is given.

## II. Simulation model

---

[1] Email: likw@ihep.ac.cn

A high intensity proton accelerator facility, which is named China Spallation Neutron Source (CSNS), is being built in China [6]. It's equipped with an H- linac and a rapid cycling synchrotron (RCS). As an example, the paper uses RCS/CSNS beam in this study. Table 1 shows the beam's parameters. The RCS beam is assumed to be cylindrical with uniform distribution in the transverse plane.

In this paper, two sources of electron cloud are (1) primary electrons at the vacuum chamber by lost protons, and (2) the multipaction from secondary electron emission. The primary electrons produced by lost protons is $Y*P_{loss}*N_p$ per turn, where $Y$ is the effective electron yield per lost proton, $P_{loss}$ is the proton loss rate per turn, and $N_p$ is the population of the bunch. The secondary electron emission yield (SEY) is defined as the fraction of the number of electrons emitted from the surface of vacuum chamber to the total number of incident electrons. When the SEY is larger than unity, the number of electrons increases exponentially. The SEY is approximated by the formula [7],

$$\delta(E) = \delta_{max} \frac{E}{E_{max}} \frac{1.44}{0.44 + (E/E_{max})^{1.44}},$$

where $E$ is the incident electron energy, $\delta_{max}$ is the largest SEY when the incident energy is $E_{max}$. Figure 1 shows the SEY for $\delta_{max} = 2.1$ and $E_{max} = 200$ eV, which is used for the simulation.

Table 1: The main parameters of the RCS/CSNS

| Parameters | Injection Phase | Extraction Phase |
| --- | --- | --- |
| Circumference $C$ (m) | 227.92 | 227.92 |
| Energy $E$ (GeV) | 0.08 | 1.6 |
| Bunch Population $N_p$ ($10^{13}$) | 1.56 | 1.56 |
| Proton loss rate per turn $P_l$ ($10^{-6}$) | 4.0 | 4.0 |
| Revolution Frequency $\omega$ (MHz) | 3.20 | 6.78 |
| Bunch length $\sigma_p$ (m) | 48.93 | 22.285 |
| Harmonic number $h$ | 2 | 2 |
| Beam transverse size $\sigma_x$, $\sigma_y$ (cm) | 1.5, 1.5 | 1.2, 1.2 |
| Pipe radius $b$ (cm) | 10 | 10 |

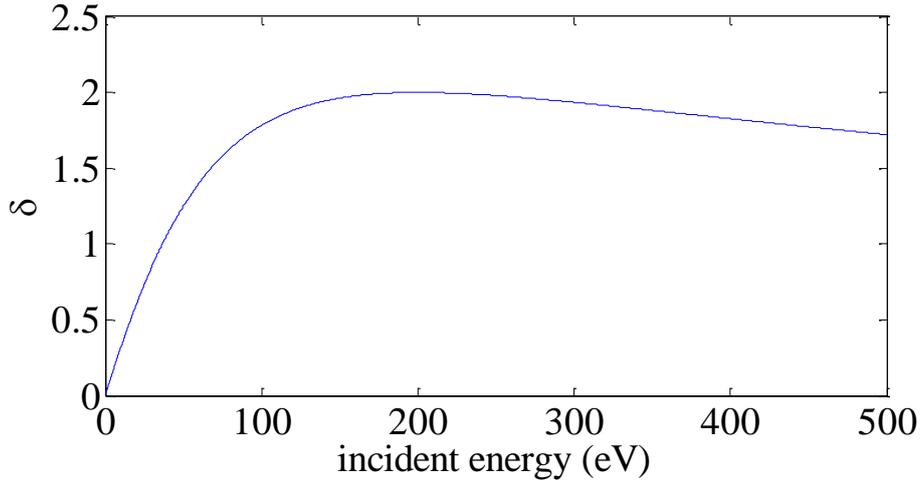

Fig. 1. Secondary emission yield with normal incident angle.

The simulation program that we used is a three-dimensional particle-in-cell (PIC) code [8]. It includes the three-dimensional electron space charge, beam-electron interaction. The primary electrons are emitted by the lost protons hitting the beam pipe when a proton bunch passes. The electrons move under the beam and its space charge. When an electron hits the surface of vacuum chamber, it generates secondary electrons. The energy and angle of secondary electrons are controlled by a statistical distribution generator obeying the experiments.

### III. Motion of electron

The electrons move under the beam electric and its own space charge. For a cylindrical beam with a uniform transverse distribution, the space-charge field is

$$E_r(t) = \begin{cases} \dfrac{\lambda(t)}{4\pi\varepsilon_0}\dfrac{2}{r} & (r > a), \\ \dfrac{\lambda(t)}{4\pi\varepsilon_0}\dfrac{2r}{a^2} & (r < a), \end{cases} \quad (1)$$

where $\varepsilon_0 = 8.85*10^{-12}$ F/m is known as the permittivity of vacuum, $\lambda$ is the beam's line density, and a is the transverse beam size. The linear oscillation frequency of the electron under the beam force is

$$f = \frac{1}{2\pi}\sqrt{\frac{r_e \lambda c^2}{a^2}}, \quad (2)$$

where $r_e$ is the classical electron radius, and $c$ is the velocity of light. For RCS beam, if the beam is longitudinal uniform, the oscillation frequency is 67.4 MHz. If the line density is bigger, the oscillation frequency will be larger. The number of oscillation under the bunch is several tens, so the bunch is called long-bunch, which is different with the bunch in electron accelerator. When a proton bunch passes an electron cloud region, the electrons are trapped under the beam because of the increasing beam line density. After the peak of bunch, the electrons' oscillation amplitude increases for the decreasing beam line density. So the electrons are released. Furthermore, the primary electron produced after the peak of bunch can move straight to the opposite wall, gain energy from the beam and generate the secondary electrons. Figure 2 shows the orbit of the electron under beam force. During the primary electron produced after the peak of bunch moving from the beam pipe to the opposite wall, the energy gain of the electron from the beam is [9]

$$\Delta E = -\frac{1}{2}\beta c \sqrt{\frac{me}{2\pi\varepsilon_0}} \frac{\partial \lambda}{\partial z} \frac{1}{\sqrt{\lambda}} \left( a(2\zeta - 1)\arcsin\frac{1}{\sqrt{\zeta}} + a\sqrt{2\ln\frac{b}{a}} \right.$$
$$\left. + \sqrt{2}\zeta \int_a^b \frac{dr}{\sqrt{\ln(b/r)}} - \frac{1}{\sqrt{2}} \int_a^b \frac{1 + 2\ln(r/a)}{\sqrt{\ln(b/r)}} dr \right), \quad (3)$$

where $\zeta = 1 + 2\ln(b/a)$, and b is the transverse of the vacuum chamber. From figure 1, when the incident energy is above about 50 eV, the SEY is above 1. So when $\Delta E$ is above 50 eV, the multipacting of the electron will happen.

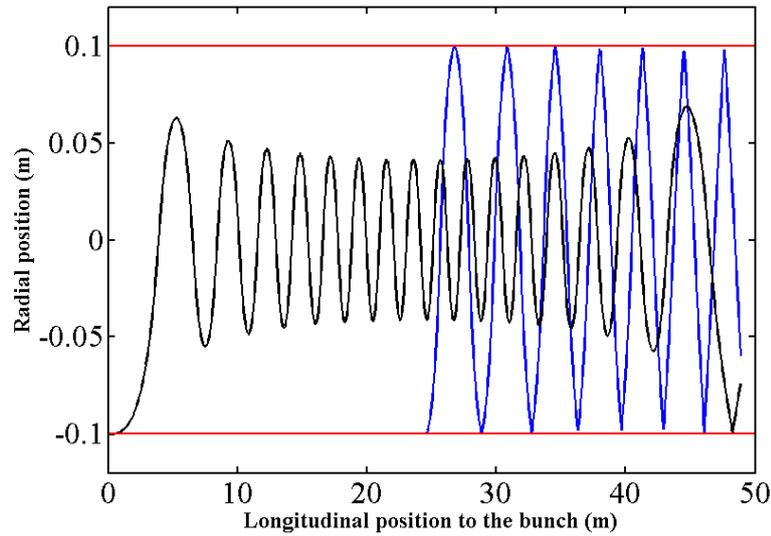

Fig. 2. The X axis is the longitudinal position of the electron to the bunch; the Y axis is the radial position of the electron in the transverse plane. The bunch is 49 m in longitudinal direction, and has a sine longitudinal distribution. The red line is the boundary of the wall; the black line is the orbit of the electron produced at the head of the bunch; the blue line is the orbit of the electron

produced after the peak of the bunch.

## IV. Simulation result

The process of the trailing-edge electron multipacting is depicted in Figure 3. The electrons are trapped at the peak of bunch. After the peak, the electrons are released, and hit the wall, generating the secondary electron emission. At the tail of the bunch, there is a strong multipacting.

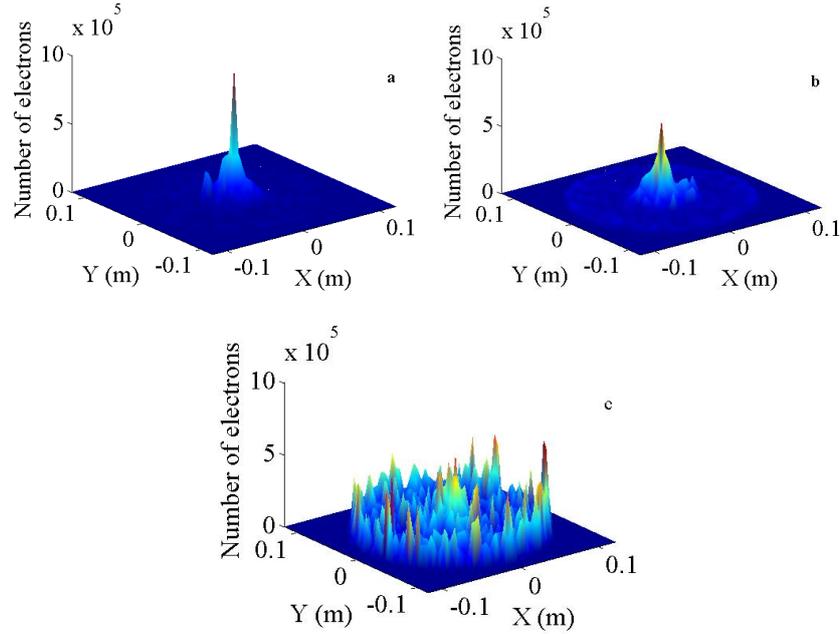

Fig. 3. The distribution of the electrons at different time: a is the distribution of the electrons at the center of the bunch; b is between the center and tail of the bunch; c is at the tail of the bunch.

We explore the electron multipacting under different longitudinal distribution beam. Figure 4 shows the electron cloud buildup under four kind of longitudinal distribution, which include gaussian, sine, elliptical, and uniform. The electron multipacting is the strongest under the gaussian longitudinal distribution beam. There is no trailing-edge electron multipacting under the uniform longitudinal distribution. It's because of no energy gain from the beam.

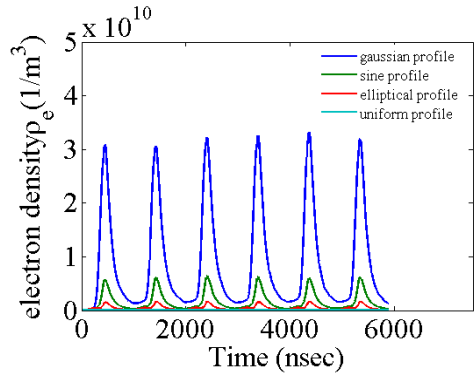

Fig. 4. Electron cloud buildup under different longitudinal distribution.

Figure 5 shows the modified RCS beam profile, compared to the sine profile. The bunch population is the same under two different beam profiles. So the primary electron produced by beam loss is equal under two different profiles. Figure 6 shows the electron cloud under the truncated tail beam profile. The density of the electron cloud decreases dramatically as the tail of the beam is progressively truncated. It's because of the breaking of the multipacting of the electron at the tail of the bunch.

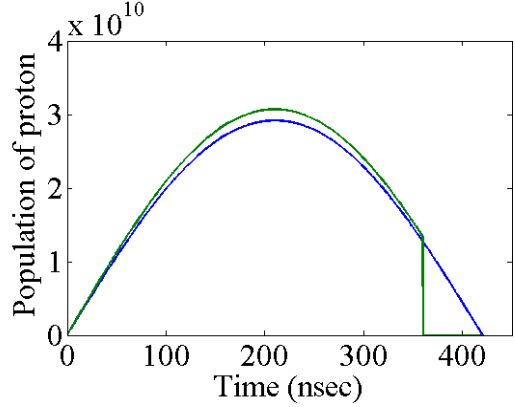

Fig. 5. Beam profile cut at 360 ns, compared to the sine profile.

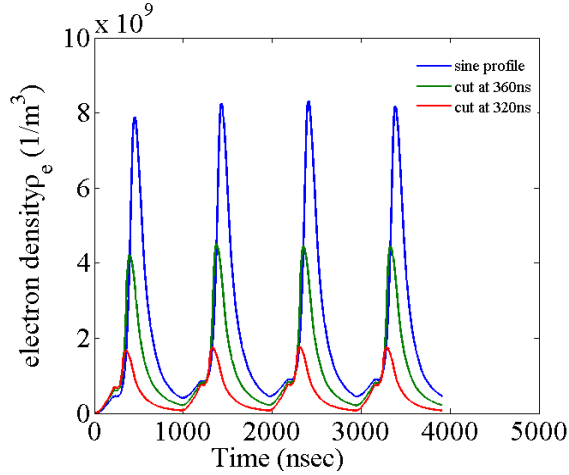

Fig. 6. The electron cloud under the beam profile which is truncated at the tail of the bunch.

## V. Conclusion

We have explored the multipacting of the electron in the long-bunch beam. The multipacting of the electron is determined by the slope of the bunch in the longitudinal direction, which releases the electron and makes the electron have enough energy gain and hit the wall, generating the secondary emission. The simulation result shows that making the bunch longitudinal profile uniform and truncating the tail of the bunch can decrease much of the electron cloud.